\documentclass[twocolumn,aps,pra,amsmath,amssymb,superscriptaddress,tightenlines]{revtex4-2}

\usepackage{amssymb}
\usepackage{amsmath}
\usepackage{graphicx}
\usepackage{color}
\usepackage{subfigure}
\usepackage{amsfonts}
\usepackage{float}     
\usepackage[colorlinks,urlcolor=blue,linkcolor=blue,citecolor=blue]{hyperref}

\def\narrowtext{\par\global\columnwidth20.5pc
	\global\hsize\columnwidth\global\linewidth\columnwidth
	\global\displaywidth\columnwidth}

\begin{document}

  \title{Near-Perfect Single-Photon Source via Ultrastrong Coupling  }
	\author{Ying Ren}
	\affiliation{Key Laboratory of Low-Dimensional Quantum Structures and Quantum Control of
		Ministry of Education, Key Laboratory for Matter Microstructure and Function of Hunan Province, Department of Physics and Synergetic Innovation Center for Quantum Effects and Applications, Hunan Normal University, Changsha 410081, China}
		\author{Ying-Xue Ma}
		\affiliation{Key Laboratory of Low-Dimensional Quantum Structures and Quantum Control of
			Ministry of Education, Key Laboratory for Matter Microstructure and Function of Hunan Province, Department of Physics and Synergetic Innovation Center for Quantum Effects and Applications, Hunan Normal University, Changsha 410081, China}
	\author{Jin-Feng Huang}
    \email{Contact author: jfhuang@hunnu.edu.cn}


	\affiliation{Key Laboratory of Low-Dimensional Quantum Structures and Quantum Control of
		Ministry of Education, Key Laboratory for Matter Microstructure and Function of Hunan Province, Department of Physics and Synergetic Innovation Center for Quantum Effects and Applications, Hunan Normal University, Changsha 410081, China}
	\affiliation{Institute of Interdisciplinary Studies, Hunan Normal University, Changsha 410081, China}
	\affiliation{Hunan Research Center of the Basic Discipline for Quantum Effects and Quantum Technologies, Hunan Normal University, Changsha 410081, China}

	\begin{abstract}

      Deterministic single-photon sources are indispensable core devices for quantum information technology, yet high-performance implementation remains a long-standing bottleneck for linear optical quantum computing. We propose a feasible scheme for deterministic single-photon emission based on a $\triangle$-type three-level atom coupled to a single-mode cavity, driven by two classical external fields, which is adaptable to both strong and ultrastrong cavity-atom coupling regimes. Under continuous-wave driving, the system achieves excellent single-photon characteristics: the normalized equal-time second-order correlation function reaches $g^{(2)}(0)\sim10^{-6}$, with a photon indistinguishability of $98.73\%$ and a state purity of $99.95\%$ in the strong coupling regime, while the ultrastrong coupling regime further suppresses $G^{(2)}(0)\sim10^{-8}$, yielding an indistinguishability of $99.10\%$ and a purity of $99.99\%$. For pulsed driving in the ultrastrong coupling regime, the source realizes superior performance, with an emission efficiency, indistinguishability, and purity of $99.96\%$, $98.98\%$, and $99.99\%$ under resonant conditions, and $100\%$, $95.91\%$, and $99.93\%$ under detuned conditions, respectively. The near-ideal optical performance of the proposed scheme provides a viable route for constructing high-quality deterministic single-photon sources, which offers a promising solution to the limitations of conventional single-photon devices and facilitates the further development of quantum information science and fundamental quantum optical research.

	\end{abstract}
	\date{\today}
	\maketitle	
	\narrowtext

	\section{Introduction}~\label{introduction}	
Single-photon sources are crucial for quantum information science \cite{science}, ultrasensitive quantum sensing and metrology \cite{sensing,metrology}, quantum technologies \cite{technology}, and fundamental quantum physics \cite{2}. In particular, they serve as key hardware for quantum communication \cite{2,Quantum communication-1,Quantum communication-2,Quantum communication-3}, photonic quantum computing \cite{3,JWPan2025np,Quantum computing-1,Quantum computing-2}, and quantum networks \cite{Quantum networks-1,Quantum networks-2}. An ideal single-photon source is capable of deterministically emitting one photon per excitation, and is expected to simultaneously exhibit near-unity purity (\(g^{(2)}(0)\rightarrow0\)), high photon indistinguishability ($I\rightarrow1$), unit efficiency, scalability, and other favorable performances. Nevertheless, realizing a source that satisfies all these requirements remains a formidable challenge, and trade-offs among these figures of merit are generally unavoidable. A low second-order correlation $g^{(2)}(0)<0.01$ is difficult to attain due to multiphoton emission and environmental noise. Photon indistinguishability is easily degraded by spectral broadening of emitted photons and coupling fluctuations in circuit cavity quantum electrodynamics. Additionally, the overall efficiency is restricted by mode mismatch between emitters such as quantum dots and nitrogen-vacancy (NV) centers and optical cavities, interfacial losses, as well as the intrinsic limits imposed by the cavity quality factor Q and mode volume V.

 Single-photon sources based on individual quantum emitters such as atoms \cite{atoms,rydberg}, molecules \cite{molecules}, ions \cite{ions}, NV centers \cite{color}, 2D materials \cite{2D}, quantum dots \cite{dots-1,dots-2} and cavity QED systems \cite{QED-1} have been realized experimentally. Semiconductor quantum dots are currently the most mature platform, but they are restricted by fixed emission wavelengths and limited array scalability. A recent experiment on a microcavity-coupled quantum dot at 4 K reports a system efficiency of 71.2\%, photon indistinguishability of 98.6\% and single-photon purity over 98\% \cite{JWPan2025np}. Nevertheless, high-performance single-photon sources are still highly desirable for theoretical and experimental research \cite{merit}.

We put forward a practical scheme to generate single photons deterministically. We study a \(\Delta\)-type three-level atom~\cite{chiral-1,chiral-2,artificial-2,artifical symmetry-broken-1,artifical symmetry-broken-2} confined in a high-finesse cavity. Its two excited levels couple to a single cavity mode to form the Jaynes-Cummings (JC) model \cite{JC0,JC-1,JC-2,JC-3}. Two classical fields resonantly drive the transitions from the ground state to the two excited states. Under resonant driving, the system reduces to an effective \(\Lambda\)-type three-level system. A Raman transition via the lowest single-excitation eigenstate of the JC model gives rise to single-photon emission. For large detuning, the \(\Lambda\) system can be further simplified into a two-level system for deterministic photon generation. We explore continuous-wave and pulsed driving for strong and ultrastrong coupling regimes. For resonant continuous-wave driving in the strong coupling regime, we achieve \(g^{(2)}(0)\sim10^{-6}\), 98.73\% indistinguishability and 99.95\% purity. In the ultrastrong coupling regime, \(G^{(2)}(0)\sim10^{-8}\), with indistinguishability of 99.10\% and purity of 99.99\%. Pulsed driving also delivers outstanding performance: in the ultrastrong coupling regime, the efficiency, indistinguishability and purity reach 99.96\%, 98.98\% and 100\% at resonance, and 100\%, 95.91\% and 99.93\% off resonance. All metrics meet the criteria of linear optical quantum computing without performance trade-offs, demonstrating near-ideal single-photon emission. Our scheme supports wavelength tuning via cavity mode frequency and works across a broad parameter space, fully compatible with current experimental capabilities. It can effectively overcome the wavelength limitation of quantum dot systems. 

The rest of this paper is organized as follows. Section \ref{model} presents the physical model and Hamiltonian, and analyzes the system dynamics under resonant and detuned conditions. Section \ref{strong coupling} investigates the system performance with dissipative effects and compares the performance of different driving strategies in the strong coupling regime. Section \ref{ultrastrong} further explores the dissipative system performance and conducts a comparative analysis of diverse driving schemes in the ultrastrong coupling regime. Conclusions and experimental outlook are given in Section \ref{conclusion}.

	\section{Model and Hamiltonian}~\label{model}
	\begin{figure}[tbp]
		\centering
		\includegraphics[width=0.48\textwidth]{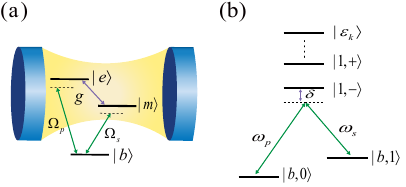}
		\caption{(Color online) (a) Schematic of the system. A  $\triangle$-type three-level atom interacts with the cavity mode via its two excited states \(|e\rangle\) and \(|m\rangle\). The transitions \(|m\rangle\leftrightarrow|b\rangle\) and \(|e\rangle\leftrightarrow|b\rangle\) are driven by two external fields with frequencies \(\omega_p\) and \(\omega_s\), respectively. The excited states \(|e\rangle\) and \(|m\rangle\) are resonantly coupled to a single cavity mode. (b) Energy-level diagram of the effective \(\Lambda\) system. The initial state \(|b,0\rangle\) and final state \(|b,1\rangle\) are connected by the driving field through the eigenstates \(|\varepsilon_k\rangle\) of the Jaynes-Cummings model. The detuning \(\delta\) is defined as the difference between the driving frequency \(\omega_p\) (\(\omega_s\)) and the transition frequency from \(|b,0\rangle\) (\(|b,1\rangle\)) to \(|\varepsilon_k\rangle\).}
		\label{Fig1_model}
	\end{figure}		

We consider a $\Delta$-type three-level atomic system, where the two excited energy levels $|e\rangle$ and $|m\rangle$ are resonantly coupled to a single-mode cavity field with resonant frequency $\omega_{c}$. The energy spacing between the intermediate level $|m\rangle$ and the ground level $|b\rangle$ is set to be much larger than the cavity frequency $\omega_{c}$, such that the atom residing in the ground state $|b\rangle$ does not couple to the cavity field. The system is initially prepared in the state $|\psi(0)\rangle=|b,0\rangle$, corresponding to the atom in the ground state $|b\rangle$ and the cavity in the vacuum state. To achieve deterministic single-photon emission, two external classical driving fields with frequencies $\omega_{s}$ and $\omega_{p}$ are applied to drive the atomic transitions $|e\rangle\leftrightarrow|b\rangle$ and $|m\rangle\leftrightarrow|b\rangle$, respectively, as schematically illustrated in Fig.~\ref{Fig1_model}(a). The total Hamiltonian of the system reads
	 \begin{eqnarray}
	 H=H_{0}+H_{1},
	\end{eqnarray}
 where $H_{0}$ represents the part without driving fields ($\hbar=1$):
	\begin{eqnarray}
		H_{0} & = & \omega_{c}a^{\dagger}a+\omega_{e}|e\rangle\langle e|+\omega_{m}|m\rangle\langle m|+\omega_{b}|b\rangle\langle b|\nonumber \\
		&  & +g(|e\rangle\langle m|a+a^{\dagger}|m\rangle\langle e|),\label{eq:1}
	\end{eqnarray}
and $H_{1}$ describes the interaction part with the two external driving fields:
	 \begin{eqnarray}
	 	H_{1} & = & \Omega_{p}\cos(\omega_{p}t)(|e\rangle\langle b|+|b\rangle\langle e|)\nonumber \\
	 	&  & +\Omega_{s}\cos(\omega_{s}t)(|m\rangle\langle b|+|b\rangle\langle m|).\label{eq:2}
	 \end{eqnarray}
Here, $a(a^{\dagger})$ is the annihilation (creation) operator of the cavity field, $\omega_{i}(i=e,m,b)$ is atomic frequency of the atomic level $|i\rangle$, $g$ is the atom-cavity coupling strength, and $\Omega_{p}$ and $\Omega_{s}$ are the Rabi frequencies of the two external driving fields.
		
Note that $H_{JC}=\omega_{c}a^{\dagger}a(|e\rangle\langle e|+|m\rangle\langle m|)+\omega_{e}|e\rangle\langle e|+\omega_{m}|m\rangle\langle m|+g(|e\rangle\langle m|a+a^{\dagger}|m\rangle\langle e|)$ describing the JC model~\cite{JC0}, in terms of the  eigenstates $|\varepsilon_{k}\rangle$ of $H_{JC}$ satisfying $H_{JC}|\varepsilon_{k}\rangle=\varepsilon_{k}|\varepsilon_{k}\rangle$, we can rewrite $H_{0}=H_{JC}+\omega_{c}a^{\dagger}a|b\rangle\langle b|+\omega_{b}|b\rangle\langle b|$, as diagonal form:
	\begin{equation}	H_{0}=\sum_{k=0}^{\infty}\varepsilon_{k}|\varepsilon_{k}\rangle\langle\varepsilon_{k}|+\sum_{n=0}^{\infty}(\omega_{b}+n\omega_{c})|b,n\rangle\langle b,n|.\label{eq:3}
	\end{equation} 	
The states $|\varepsilon_{k}\rangle$ are defined as $|\varepsilon_{0}\rangle\equiv|m,0\rangle$, $|\varepsilon_{2n-1}\rangle\equiv|n,-\rangle =-\sin\theta_{n}|e,n-1\rangle+\cos\theta_{n}|m,n\rangle$, and $|\varepsilon_{2n}\rangle\equiv|n,+\rangle=\cos\theta_{n}|e,n-1\rangle+\sin\theta_{n}|m,n\rangle(n=1,2\ldots)$. Here, the subscript $n$ label the $n$ excitation subspace of the JC Hamiltonian and the angle $\theta_{n}$ are defined by $\cos\theta_{n}=\left(2g\sqrt{n}\right)/[(\tilde{\Omega}_{n}-\Delta)^{2}+4g^{2}n]^{1/2}$, where $\tilde{\Omega}_{n}=\sqrt{\Delta^{2}+4g^{2}n}$ is the generalized Rabi frequency and $\Delta=\omega_{e}-\omega_{m}-\omega_{c}$ is the detuning from the cavity field of the level space between $|e\rangle$ and $|m\rangle$. Meanwhile, we define the probability amplitudes as	$c_{kn}=\langle \varepsilon_{k}|m,n\rangle$ and $d_{kn}=\langle \varepsilon_{k}|e,n\rangle$. Similarly, the driving Hamiltonian $H_{1}$ can be rewritten as:
   	\begin{equation}
   		H_{1}=\sum_{n,k=0}^{\infty}\lambda_{kn}(t)\left(|b,n\rangle\langle \varepsilon_{k}|+\varepsilon_{k}\rangle\langle b,n|\right),\label{eq:4}
   	\end{equation}
where  $\lambda_{kn}(t)\equiv\Omega_{p}d_{kn}\cos(\omega_{p}t)+\Omega_{s}c_{kn}\cos(\omega_{s}t)$ is introduced.

$H_{1}$ describes many transition processes, where certain resonant transitions can be selected by tuning the driving field frequencies $\omega_{p}$ and  $\omega_{s}$. We focus on the Raman resonant condition
   	\begin{equation}
   		\omega_{p}-\omega_{s}=\omega_{c},\label{eq:5}
   	\end{equation}
    which enables resonant transitions from the initial state $|b,0\rangle$ to the target state $|b,1\rangle$ via intermediate state $|1,-\rangle$	and some higher-energy levels through the Raman transition, as shown in Fig.~\ref{Fig1_model}(b). When the driving fields is sufficiently weak, the resonant coupling is dominant and only the states involved in the resonant coupling can be effectively accessed. 
   	
In the interaction picture defined by  $U(t)=\exp(-iH_{0}t)$, the system Hamiltonian $H$ becomes
   	\begin{eqnarray}
   		H_{I} & = & \sum_{n,k=0}^{\infty}\sum_{q=\pm1}\left(\chi_{p,kn}e^{i\delta_{kn,qp}t}+\chi_{s,kn}e^{i\delta_{kn,qs}t}\right)|\varepsilon_{k}\rangle\langle b,n|\nonumber \\
   		&  & +\textrm{H.c.},\label{eq:6}
   	\end{eqnarray}
where $\chi_{p,kn}$, $\chi_{s,kn}$ and the detuning $\delta_{kn,ql}~(l=p,s)$ are defined as
   	\begin{eqnarray}
   		\chi_{p,kn} & = & \frac{\Omega_{p}}{2}d_{kn},\nonumber \\
   		\chi_{s,kn} & = & \frac{\Omega_{s}}{2}c_{kn},\nonumber \\
   		\delta_{kn,ql} & = & \varepsilon_{k}-\omega_{b}-n\omega_{c}+q\omega_{l},~(q=\pm1).\label{eq:7}
   	\end{eqnarray}
   	The effective coupling strength $\chi_{p,kn}$ and $\chi_{s,kn}$ are dependent on the coefficient $d_{kn}$ and $c_{kn}$. When the cavity field resonatly couple to with the upper two levels $|e\rangle$ and $|m\rangle$, namely $\Delta=0$, $d_{kn}$ and $c_{kn}$ become constants. As we shall discuss in the following $H_{I}$ can be well approximated by keeping only $n=0$ and $n=1$ terms.

\subsection{Large detuning case}   
	We first consider the large detuning case defined by the 
conditions: $\delta\equiv\varepsilon_{1}-\omega_{b}-\omega_{p}\gg (\Omega_{p}\cos\theta_{1}/2)$
and $\varepsilon_{1}-\omega_{b}-\omega_{c}-\omega_{s}\gg (\Omega_{s}\sin\theta_{1}/2)$ under which the energy levels $|\varepsilon_{k}\rangle(k=0,1,2,...)$ can be adiabatically eliminates.
Using the time-averaging method~\cite{time-averaging procedure-1,time-averaging procedure-2} with Raman resonance condition (\ref{eq:5}), we can adiabatically eliminate
the energy levels $|\varepsilon_{k}\rangle(k=0,1,2,...)$, then the Hamiltonian
$H_{I}$ becomes:
\begin{eqnarray}
	H'_{eff} & \cong & \sum_{n=0}^{\infty}\Delta_{n}|b,n\rangle\langle b,n|+\sum_{n=0}^{\infty}g_{n,n+1}|b,n\rangle\langle b,n+1|\nonumber \\
	&  & +\textrm{H.c.}.\label{eq:A4}
\end{eqnarray}
Here the ac Stark shifts $\Delta_{n}$ and effective coupling strength
$g_{n,n+1}$ are defined as
\begin{eqnarray}
	\Delta_{n} & = & -\sum_{k=0}^{\infty}\sum_{l=s,p}\sum_{q=\pm1}\frac{|\chi_{kn,l}|^{2}}{\delta_{kn,ql}},\label{eq:A5}\\
	g_{n,n+1} & = & 
	-\sum_{k=0}^{\infty}\frac{\left(\Omega_{s}c_{kn+1}\right)\left(\Omega_{p}d_{kn}\right)^{*}}{4\delta_{kn,-1p}}.\label{eq:A6}
\end{eqnarray}

During the derivation of Eq.~(\ref{eq:A4}), we have discard the rapidly oscillating terms, such as higher-order transition
terms of the form $|b,n+m\rangle\langle b,n|(m\ge2)$. This approximation is valid due to the detuning difference in the multiphoton process $|b,n\rangle\rightarrow|\varepsilon_{k}\rangle\rightarrow|b,n+m\rangle$
with $m\geqslant2$, is significantly
larger than that in the single-photon Raman process $|b,n\rangle\rightarrow|\varepsilon_{k}\rangle\rightarrow|b,n+1\rangle$.

Under the condition of weak driving with $\Omega_{p}/ \omega_{c}<0.01$, we find that
$g_{1,2}\ll g_{0,1}$. With the condition $g_{1,2}t\ll1$, the transition from $|b,1\rangle$ to
$|b,2\rangle$ can be ignored. Eq.~(\ref{eq:A4}) is therefore further
simplified, enabling a valid truncate and yielding the effective
Hamiltonian
\begin{eqnarray}
	H''_{eff} & = & \Delta_{0}|b,0\rangle\langle b,0|+\Delta_{1}|b,1\rangle\langle b,1|\nonumber \\
	&  & +g_{0,1}|b,0\rangle\langle b,1|+\textrm{H.c.}.\label{eq:A7}
\end{eqnarray}	 
Thus, the system simplifies to an effective two-level system, where coherent Rabi oscillations take place between the states $|b,0\rangle$ and $|b,1\rangle$. Specifically,
when the detuning satisfies $\Delta_{0}=\Delta_{1}$, at time
$t=\pi/(2g_{0,1})$, the state $|b,0\rangle$ is completely
transfered to the state $|b,1\rangle$, generating a single photon due to $|b\rangle$ dose not couple to the cavity. The single cavity photon is free as a real photon, when the driving field is switched off at time $t=\pi/(2g_{0,1})$. Physically, the creation of the single photon is due to the intermediate state $|1,-\rangle$ where $\sin\theta_{1}|m,1\rangle$ provides the transition matrix element.

We note that the condition $\Delta_{0}=\Delta_{1}$ corresponds to
the balanced state of ac Stark shifts. This condition can be achieved
by precisely tuning the intensity ratio of the driving fields at the ratio 
\begin{equation}
	\alpha\equiv\frac{\Omega_{s}}{\Omega_{p}}=\sqrt{-\frac{f(d_{k0},d_{k1},\omega_{p})}{f(c_{k0},c_{k1},\omega_{s})}}
\end{equation}
where
\begin{eqnarray}
	f(x,y,z) & = & \sum^{\infty}_{k=0}\frac{|x|^{2}(\varepsilon_{k}-\omega_{b})}{(\varepsilon_{k}-\omega_{b})^{2}-z^{2}}\nonumber \\
	&  & -\sum^{\infty}_{k=0}\frac{|y|^{2}(\varepsilon_{k}-\omega_{b}-\omega_{c})}{(\varepsilon_{k}-\omega_{b}-\omega_{c})^{2}-z^{2}}.\label{eq:A9}
\end{eqnarray}

   	
\subsection{Resonant case}   
    Now we consider the  resonant case: the pump field resonantly drives the transition $|b,0\rangle\leftrightarrow|1,-\rangle$, i.e., $\delta=0$; and the Stokes driving fields resonantly drives the transition $|1,-\rangle\leftrightarrow|b,1\rangle$, i.e., $\varepsilon_{1}-\omega_{b}-\omega_{c}-\omega_{s}=0$. Under these conditions, all the other eigenstates $|\varepsilon_{k}\rangle(k\neq1)$ are far off resonance and can be discarded. Then Eq. (\ref{eq:6}) can be reduce to an effective $\Lambda$-type three-level Hamiltonian:
   	\begin{equation}
   		H_{eff}=\chi_{p}|\varepsilon_{1}\rangle\langle b,0|+\chi_{s}|\varepsilon_{1}\rangle\langle b,1|+\textrm{H.c.},\label{eq:A1}
   	\end{equation}
   	where $\chi_{p}=-(\sqrt{2}\Omega_{p})/4$ and $\chi_{s}=(\sqrt{2}\Omega_{s})/4$.
   	
   	By solving the Schrödinger equation with the effective Hamiltonian (\ref{eq:A1}),  we derive the single-photon transition probability from the initial state $|b,0\rangle$  to the final state $|b,1\rangle$ as
   	\begin{equation}
   		P_{1}=\frac{4\alpha^{2}|\sin\theta_{1}|^{2}|\cos\theta_{1}|^{2}}{(|\sin\theta_{1}|^{2}+\alpha^{2}|\cos\theta_{1}|^{2})^{2}}\sin^{4}\frac{\bar{\omega} t}{2}\label{eq:A2}
   	\end{equation}
   where the effective frequency is defined as $\bar{\omega}=(\Omega_{p}/2)\sqrt{|\sin\theta_{1}|^{2}+\alpha^{2}|\cos\theta_{1}|^{2}}$. When the intensity ratio  of the two driving fields reaches the optimal value 		
   	\begin{equation}
   		\alpha=\alpha_{c}=\frac{|\sin\theta_{1}|}{|\cos\theta_{1}|},\label{eq:A3}
   	\end{equation}
   	the population of state $|b,1\rangle$ achieves unity ($P_{1}=1$)at time $t={\pi}/{\bar{\omega}}$.  
   	
   	For the closed system, we have derived the effective Hamiltonian under
   	both resonant and large-detuned driving. The  exact numerical dynamics show well
   	agreement with the analytical results, which is shown in Appendix A. Moreover, we observed that
   	the population can be fully transferred from the initial state $|b,0\rangle$ to the final state $|b,1\rangle$ in either resonant or large-detuned driving, confirming the feasibility of our physical mechanism for creating single photons.

	\vspace{-2pt}
		
		
	\section{Single photons via strong coupling}~\label{strong coupling}
	
		\begin{figure}[tbp]
		\centering
		\includegraphics[width=0.48\textwidth]{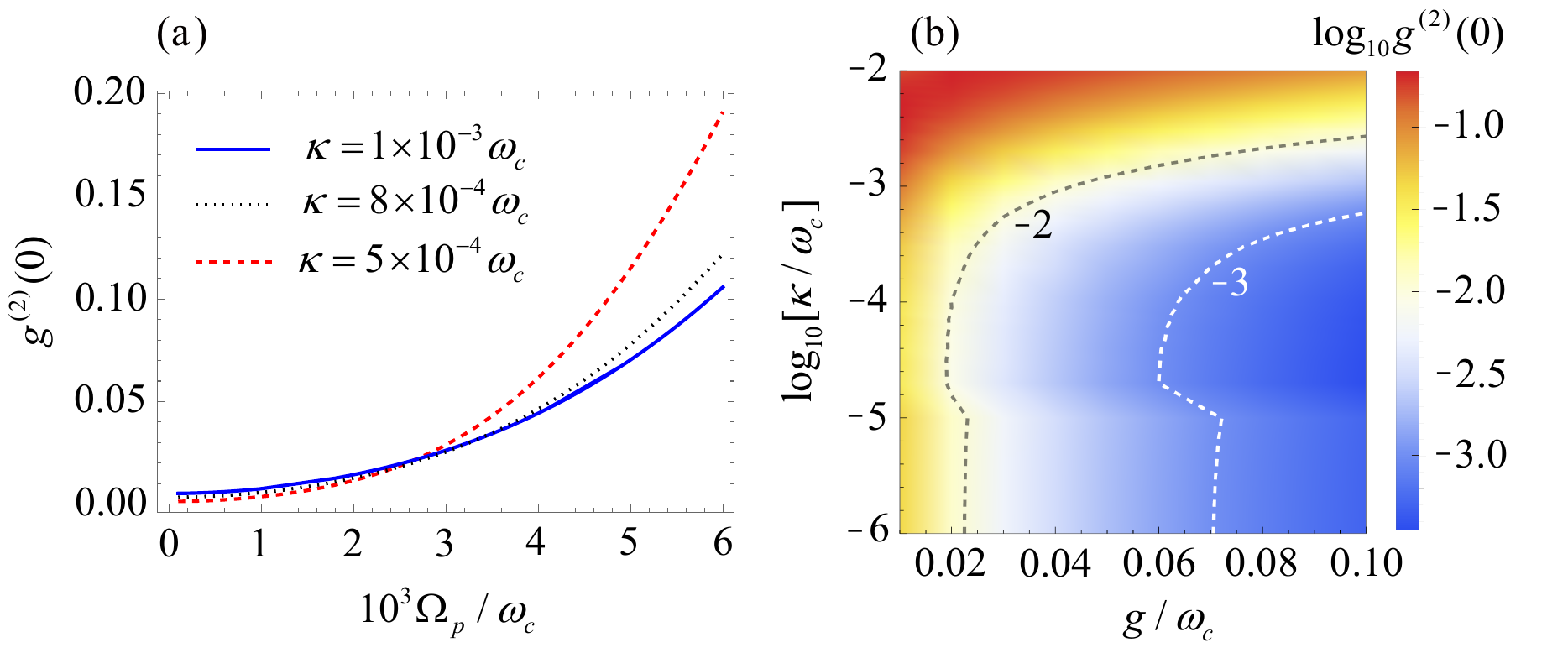}
		\caption{(Color online) (a) Equal-time second-order correlation function $g^{(2)}(0)$ as a function of the driving amplitude $\Omega_{p}$ on resonance ($\delta = 0$) with $g = 0.05\omega_{c}$ and $\kappa_{i}=\kappa = 10^{-3}\omega_{c}$ [$ (i=1,2,3)$, blue solid curve], $8\times10^{-4}\omega_{c}$ (black dotted curve), and $5\times10^{-4}\omega_{c}$ (red dashed curve), respectively.
			(b) $g^{(2)}(0)$ as functions
			of the coupling strength $g$ and the damping rate $\kappa_{i}=\kappa$ $ (i=1,2,3)$ on resonance ($\delta = 0$) for $\Omega_{p}=10^{-3}\omega_{c}$. The black and white dashed curves label $g^{(2)}(0)=10^{-2}$ and $g^{(2)}(0)=10^{-3}$, respectively. Other common 
			parameters are: $\Omega_{s}=\alpha_{c}\Omega_{p},~\omega_{s}=\omega_{p}-\omega_{c},~\omega_{b}=-5\omega_{c},~\omega_{m}=-0.5\omega_{c}$, and $\omega_{e}=0.5\omega_{c}$
			 }
		\label{Fig4-1}
	\end{figure} 
	
	In the previous section, we have investigated the generation of single photons under non-dissipative conditions. In this section, we focus on how the cavity photons can be transmitted out the cavity as real photons for $g/\omega_{c}<0.1$.
	The dynamics of the open system is governed by the master equation at zero temperature~\cite{master-equation-1} 
	\begin{equation}
		\frac{d\rho(t)}{dt}=i[\rho(t),H]+\mathcal{L}_{1}\rho+\mathcal{L}_{2}\rho+\mathcal{L}_{3}\rho,\label{eq:23-1}
	\end{equation}
	where 
	\begin{eqnarray}
		\mathcal{L}_{1}\rho & = & \kappa_{1}\left(2|m\rangle\langle e|\rho|e\rangle\langle m|-|e\rangle\langle e|\rho-\rho|e\rangle\langle e|\right),\label{eq:36-1}\notag\\
		\mathcal{L}_{2}\rho & = & \kappa_{2}\left(2|b\rangle\langle m|\rho|m\rangle\langle b|-|m\rangle\langle m|\rho-\rho|m\rangle\langle m|\right),\label{eq:36-2}\notag\\
		\mathcal{L}_{3}\rho & = & \kappa_{3}\left(2a\rho a^{\dagger}-a^{\dagger}a\rho-\rho a^{\dagger}a\right),
        \label{eq:36-3}
	\end{eqnarray}
and $\kappa_{1}$ and $\kappa_{2}$ denote the damping rate associated with the decay from  $|e\rangle$ to $|m\rangle$ and from $|m\rangle$ to $|b\rangle$ respectively, and $\kappa_{3}$ is the cavity decay rate.

	To investigate the quantum statistical behavior of the system, we calculate the time-delayed second-order correlation function~\cite{scully} 
	\begin{equation}
		g^{(2)}(\tau)=\frac{\left\langle a^{\dagger}(t)a^{\dagger}(t+\tau)a(t+\tau)a(t)\right\rangle }{\left\langle a^{\dagger}(t)a(t)\right\rangle \left\langle a^{\dagger}(t+\tau)a(t+\tau)\right\rangle }.
	\end{equation}
	
	We first investigate continuous driving case. The dependence of the equal-time second-order correlation function $g^{(2)}(0)$ on the driving amplitude $\Omega_{p}$, the coupling strength $g$ and the damping rate is shown in Fig. \ref{Fig4-1}. Here, we select the time $t$ at the first peak of the photon number $\bar{n}(t)$~\cite{peng j,Source-4}.  Fig.\ref{Fig4-1}(a) shows the dependence on $\Omega_{p}$ with  $g/\omega_{c}=0.05$ and  $\kappa_{i}/\omega_{c}=\kappa/\omega_{c}=1\times10^{-3}$, $8\times10^{-4}$ and $5\times10^{-4}$, respectively.  $g^{(2)}(0)$ increases with increasing $\Omega_{p}$ for various damping rate. Based on this phenomenon, we further investigate the dependence of $g^{(2)}(0)$ on $g$  and $\kappa$ in Fig.\ref{Fig4-1}(b) by setting $\Omega_{p}=0.001\omega_{c}$.  As shown in Fig.~\ref{Fig4-1}(b), $g^{(2)}(0)$ decreases with the increase of $g$. Meanwhile, the dependence of $ g^{(2)}(0)$ on dissipation is not monotonic due to the competition with coupling strength $g$. As indicated by the black and white dashed curves in Fig.~\ref{Fig4-1}(b), $g^{(2)}(0)$ remains on the order of $10^{-3}$ when $\kappa/\omega_{c} < 2\times10^{-3}$ and $0.025 < g/\omega_{c} < 0.1$, and even reaches the order of $10^{-4}$ when $\kappa/\omega_{c} < 6\times10^{-4}$ and $0.07 < g/\omega_{c} < 0.10$. For example, at $\kappa/\omega_{c} = 10^{-6}$ and $g/\omega_{c} = 0.08$, $g^{(2)}(0) = 5.41\times10^{-4}$. After normalizing the maximum value of $g^{(2)}(\tau)$, we obtain $g^{(2)}(0) = 5.27\times10^{-6}$, which is four orders of magnitude smaller than the result reported in Ref. \cite{Source-4,Source-5}. This demonstrates a significantly strong antibunching effect.
	
	
		\begin{figure}[t!]
		\centering
		\includegraphics[width=0.48\textwidth]{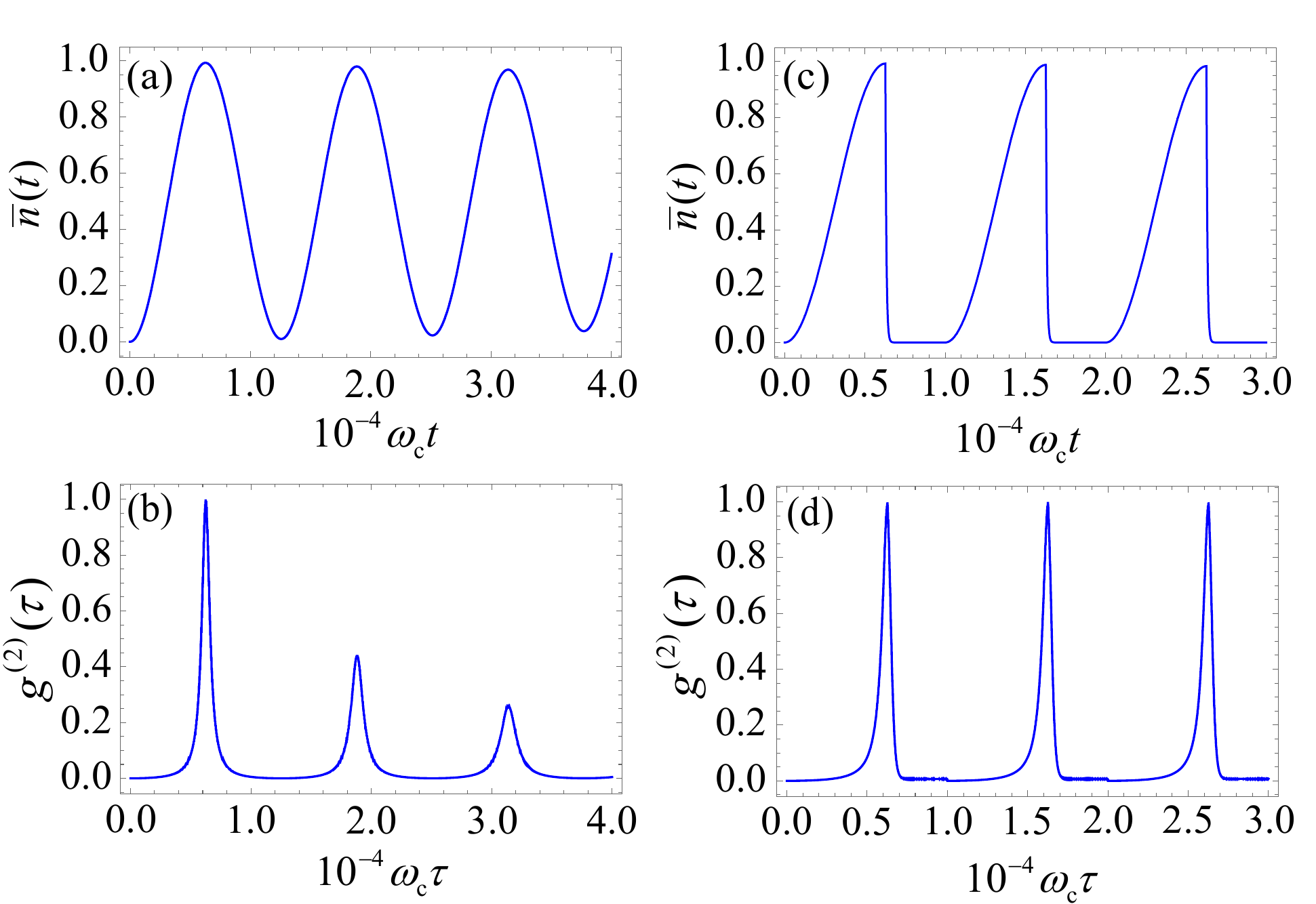}
		\caption{(Color online) (a, c) Time evolution of the average photon number $\bar{n}(t)$ under (a) continuous driving  and (c) pulse driving, respectively. (b, d) Normalized second-order 
			correlation function $g^{(2)}(\tau)$ as a function of the delay time $\tau$ under (b) continuous driving and (d) pulse driving, respectively.  
           The parameters are set as: (a, b) $\kappa_{i}=10^{-6}\omega_{c}$ ($i=1,2,3)$ and (c, d ) $T=10^{4}/\omega_{c}$, $\kappa_{0}=10^{-6}\omega_{c}$ and $\kappa'=8\times10^{-3}\omega_{c}$ 
            with $\delta=0$, $\Omega_{p}=10^{-3}\omega_{c}$  and $g=0.08\omega_{c}$.
            Other parameters are the same
			as in Fig.~\ref{Fig4-1}.
		}
		\label{Fig3-1}
	\end{figure}  
		

	To see clearly the single-photon emission dynamics, we further plot the evolution of the mean photon number $\bar{n}(t)=\left\langle a^{\dagger}(t)a(t)\right\rangle $  in Fig.~\ref{Fig3-1}(a). When  $\kappa/\omega_{c}=10^{-6}$, and  $g/\omega_{c}=0.08$, it is clearly observed that the photon number exhibits coherent oscillations between $0$ and $1$ accompanied by a slowly reducing of amplitude. Specifically, the amplitudes of the first three oscillation peaks were 0.993, 0.98, and 0.969 respectively, and eventually decayed to a stable value. This phenomenon reveals that single photons can be emitted at $t=nT$ with periodicity $T\approx1.26\times10^{-2}/\kappa$. When the system reaches a steady state, the system is in the coherent superposition of $|b,0\rangle$ and $|b,1\rangle$. This indicates that, under weak continuous driving, the system evolves into a dynamically balanced steady state. This confirms that the two or more photons are completely suppressed.  To further veryify the analysis, we plot the normalized $ g^{(2)}(\tau) $  in Fig.~\ref{Fig3-1}(b) using the same parameters as in Fig.~\ref{Fig3-1}(a). Here we normalize  $ g^{(2)}(\tau) $ by the maximum amplitude of its peak.
    It is found that $g^{(2)}(0) = 5.27\times10^{-6}$ ,  and  $ g^{(2)}(\tau) $ exhibit multi peaks with the same periodicity with $\bar{n}(t)$ and height reducing apparently.  The locations of each peaks of $g^{(2)}(\tau)$ matches well with that of $\bar{n}(t)$. These peaks denote individual single-photon emission events. We point out that, the probability on multi-photons are not strongly suppressed for the off-resonance driving in a long time due to two- or more-photon processes involved for $g/\omega_{c} < 0.1$.
	
    Then, we explore the indistinguishability $I$ of the single-photon source,
which is a fundamental parameter characterizing whether quantum interference between photons can occur. 
To quantify indistinguishability, we analyze the delayed second-order correlation function $g^{(2)}(\tau)$. 
Specifically, we extract the area of the central peak around $\tau=0$, divide it by the area of the unattenuated side peaks, and subtract this ratio from 1~\cite{I}. A value closer to 1 indicates higher photon indistinguishability, reflecting better single-photon source quality. As shown in Fig.~\ref{Fig3-1}(b),  $I=98.73\%$, which approaches $100\%$. Another metric is the purity $P$, which is defined as~\cite{Source-2}
\begin{eqnarray}
P=1-g^{(2)}(0).
\end{eqnarray}
The corresponding purity $P=99.95\%$ in Fig.~\ref{Fig3-1}(b).
	
    To better generate single photons, we apply pulse driving instead of continuous driving. Specifically, we turn on and off the driving fields and varying the cavity damping rate periodically, which follows the forms
\begin{eqnarray}
\Omega_{i}(t) & = & \Omega_{i}\sum_{n=0}^{\infty}\left[\Theta(t-nT)-\Theta(t-nT-t_{1})\right],(i=p,s),\nonumber \\
\end{eqnarray}
\begin{eqnarray}
\kappa_{3}(t) & = & \sum_{n=0}^{\infty}\kappa_{0}\left[\Theta(t-nT)-\Theta(t-nT-t_{1})\right]+\nonumber \\
 &  & \sum_{n=0}^{\infty}\kappa'\left[\Theta(t-nT-t_{1})-\Theta(t-nT-T)\right],\notag\\
 \label{kt}
\end{eqnarray}
where $\Theta$ is the Heaviside step function. $T$ is the periodicity of one emission cycle. To be different from continuous driving, we change the notations $\Omega_{i}\rightarrow\Omega_{i}(t)$ and $\kappa_3\rightarrow\kappa_{3}(t)$. For strong coupling,  $t_{1}\approx\pi/(2 g_{0,1})$ (detuned) and $\pi/\bar{\omega}$ (resonant) is the moment of the first peak of  $\bar{n}(t) $. At time $t=t_{1}$, the driving fields are first turned off and the cavity decay rate $\kappa_{3}$ is simultaneously adjusted from $\kappa_{0}$ to a higher value $\kappa'$ to let single photons emit quickly~\cite{kappa}. After photons emitted, the system returns back to the ground state. Then the driving fields are turned on again at time $t=T$, beginning the second cycle with the above steps repeated.  The above single-photon generation and emission processes are periodically repeated. We numerically simulate  $\bar{n}(t)$ and $g^{(2)}(\tau)$ for this pulse driving scheme in Fig.~\ref{Fig3-1}(c) and \ref{Fig3-1}(d), respectively. 
It is observed that $\bar{n}(t)$ oscillates periodically between $0$ and $1$, with  $T = 10^4 / \omega_c$. Meanwhile, $g^{(2)}(\tau)$ shows coherent oscillations with the same period as $\bar{n}(t)$, which is a clear signature of single-photon source. Importantly, the height of all the peaks becomes the same. The corresponding single-photon efficiency~\cite{peng j} 
    \begin{eqnarray}
    \varepsilon=\kappa_3 \int \langle a^\dagger(t) a(t) \rangle dt, 
    \end{eqnarray}
is 50\%, the indistinguishability $I$ is 98.47\%, and the purity $P$ remains consistent with that under continuous driving, i.e., $P = 99.95\%$.

	\section{Single photons via ultrastrong coupling}~\label{ultrastrong}			

In this section, we further explore the ultrastrong coupling regime where \(g/\omega_{c}>0.1\). Under the conditions of weak system–bath coupling, short bath correlation time and the conventional Born–Markov approximation \cite{master-equation-1}, the dynamics of the open quantum system in the ultrastrong coupling regime is governed by the following master equation~\cite{huang,master}:
  \begin{eqnarray}
    	\frac{d\rho(t)}{dt} & = & i[\rho(t),H]\nonumber \\
    	&  & +\sum_{i=1}^{3}\sum_{n,m>n}\Gamma_{mn}^{(i)}\{\mathcal{D}[|E_{n}\rangle\langle E_{m}|]\rho(t)\}\label{eq:20}
    \end{eqnarray}
The dissipative superoperator is defined as \(\mathcal{D}[O]\rho=O\rho O^{\dagger}-\frac{1}{2}O^{\dagger}O\rho-\frac{1}{2}\rho O^{\dagger}O\). Here, \(\{E_{n}\}=\{|b,0\rangle,|b,1\rangle,\dots,|\varepsilon_{0}\rangle,|\varepsilon_{1}\rangle,\dots\}\) are the eigenstates of \(H_0\), satisfying \(H_{0}|E_{n}\rangle=E_{n}|E_{n}\rangle\). The index \(i=1,2,3\) labels three independent dissipation channels: \(i=1\) represents atomic decay from the excited state \(|e\rangle\) to the intermediate state \(|m\rangle\), \(i=2\) describes the decay from \(|m\rangle\) to the ground state \(|b\rangle\), and \(i=3\) accounts for cavity photon damping induced by transmission losses.The relaxation rate \(\Gamma_{mn}^{(i)}\) takes the form
 \begin{equation}
    	\Gamma_{mn}^{(i)}=2\pi\varrho^{(i)}\left(\omega_{nm}\right)\lambda_{j}^{(i)2}(\omega_{nm})|C_{nm}^{(i)}|^{2}.\label{eq:22}
    \end{equation} 
In the above expression, \(\omega_{nm}=E_{n}-E_{m}\) is the transition frequency between eigenstates \(|E_{n}\rangle\) and \(|E_{m}\rangle\), \(\varrho^{(i)}(\omega_{nm})\) is the spectral density of the i-th reservoir at frequency \(\omega_{nm}\), \(\lambda^{(i)}(\omega_{nm})\) characterizes the system–bath coupling strength, and $C_{n,m}^{(i)}$ denotes the corresponding transition matrix element. The explicit forms of $C_{n,m}^{(i)}$ are presented below:
    \begin{align}
   	C_{nm}^{(1)} & =\langle E_{n}|\left(|e\rangle\langle m|+|m\rangle\langle e|\right)|E_{m}\rangle,\label{eq:23}\\
    	C_{nm}^{(2)} & =\langle E_{n}|\left(|m\rangle\langle b|+|b\rangle\langle m|\right)|E_{m}\rangle,\label{eq:24}\\
    	C_{nm}^{(3)} & =\langle E_{n}|\left(a+a^{\dagger}\right)|E_{m}\rangle.\label{eq:25}
    \end{align}

      \begin{figure}[tbp]
     	\centering
     	\includegraphics[width=0.48\textwidth]{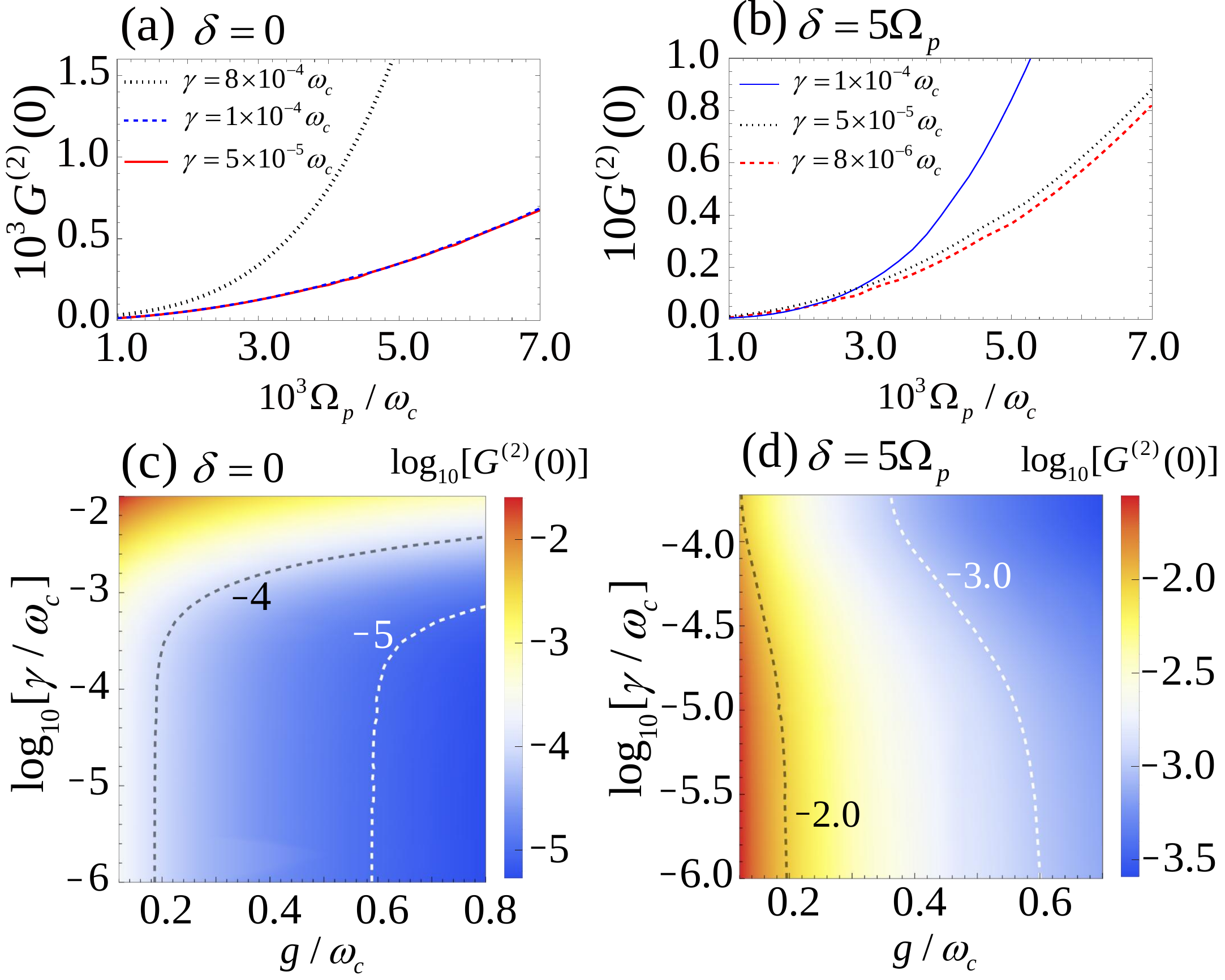}
     	\caption{(Color online) (a, b) Equal-time second-order correlation function $G^{(2)}(0)$ as
     		a function of the driving amplitude $\Omega_{p}$ with $g=0.5\omega_{c}$ for (a) resonant case ($\delta=0$) at $\gamma_{i}=\gamma=8\times10^{-4}\omega_{c}~(i=1,2,3)$ (black dotted curve), $1\times10^{-4}\omega_{c}$ (blue dashed curve) and $5\times10^{-5}\omega_{c}$ (red solid curve),  and
     		(b) detuned case ($\delta=5\Omega_{p}$) at $\gamma=1\times10^{-4}\omega_{c}$ (blue solid curve), $5\times10^{-5}\omega_{c}$ (red dashed curve),  and $8\times10^{-6}\omega_{c}$ (black dotted curve).
     		 (c, d) $G^{(2)}(0)$ vs the coupling strength $g$ and the decay rate $\gamma$ for $\Omega_{p}=1\times10^{-3}\omega_{c}$ at
     		(c) resonant case ($\delta=0$) and (d) detuned case ($\delta=5\Omega_{p}$). 
            The white and black dashed curves in (c) correspond to $G^{(2)}(0)=10^{-5}$ and  $G^{(2)}(0)=10^{-4}$, and  those in (d) correspond to $G^{(2)}(0)=10^{-2}$ and  $G^{(2)}(0)=10^{-3}$.     Other common parameters are: $\Omega_{s}=\alpha \Omega_{p}$,~ $\omega_{s}=\omega_{p}-\omega_{c},~\omega_{b}=-5\omega_{c},~\omega_{m}=-0.5\omega_{c}$ and $\omega_{e}=0.5\omega_{c}$.
     	}
     	\label{Fig5}
     \end{figure}

    In the derivation of Eq.~(\ref{eq:20}), we neglect the Lamb-shift induced corrections to the system energy levels. To simplify
    the analysis, we assume that the coupling strength $\lambda^{(i)}(\omega_{nm})$ and the spectral
    density of the environment $\varrho^{(i)}\left(\omega_{nm}\right)$
    are constants within the relevant frequency range. Under this assumption,
    the relaxation coefficient can be simplified as $\Gamma_{mn}^{(i)}=\gamma_{i}|C_{nm}^{(i)}|^{2} $.

       To study the statistical properties of the emitted photons, we need to employ the input-output theory in the ultrastrong coupling regime\cite{huang,X}. Then, we numerically examine the second-order correlation function defined by
      \begin{equation}
      	G^{(2)}(\tau)=\frac{\left\langle X^{+}(t)X^{+}(t+\tau)X^{-}(t+\tau)X^{-}(t)\right\rangle }{\left\langle X^{+}(t)X^{-}(t)\right\rangle \left\langle X^{+}(t+\tau)X^{-}(t+\tau)\right\rangle }.\label{eq:32}
      \end{equation}
      The cavity mode operator $a(a^{\dagger})$ has been replaced by the operator $X^{-}(X^{+})$ defined as 
      	\begin{eqnarray}
      	X^{-} & = & \sum_{m,n<m}C_{n,m}^{(3)}|E_{n}\rangle\langle E_{m}|,\label{eq:29}\\
      	X^{+} & = & (X^{-})^{\dagger},\label{eq:30}
      \end{eqnarray}
      which describes the transition between the eigenstates of the ultrastrongly coupled system. To be practical, we select time $t$ in $G^{(2)}(\tau)$ the moment when
      the photon number $\langle X^{+} X^{-} \rangle$ reaches its first maximum,  namely when the first photon is emitted.
      
     We first study the continous driving scheme. We investigated the dependence of $G^{(2)}(0)$ on the driving amplitude $\Omega_{p}$ in both resonant and detuned cases at $g/\omega_{c}=0.5$, and show the results in Figs.~\ref{Fig5}(a) and \ref{Fig5}(b). It can be observed that, whether in the resonant case ($\delta = 0$) or the detuned case ($\delta = 5\Omega_p$), $G^{(2)}(0)$ monotonically increases with the increase of $\Omega_{p}$ under various decay rates.   With the guidance of the above result, we further explore the dependence of $G^{(2)}(0)$ on the coupling strength $g$ and the decay rate $\gamma_{i}=\gamma~(i=1,2,3)$ at $\Omega_{p}=1\times10^{-3}\omega_{c}$ and show the results in Figs.~\ref{Fig5}(c) and \ref{Fig5}(d) for resonant and detuned cases,  respectively. Observation reveals that in both resonant and detuned cases, $G^{(2)}(0)$ decreases as the coupling strength $g$ increases, but it has a non-monotonic dependence on the decay rate $\gamma$. This non-monotonicity stems from the competition between $g$ and $\gamma$. In both the resonant and detuned cases, as the dissipation $\gamma$ increases, the coupling between the cavity mode and the external thermal reservoir is strengthened, leading to a shortened cavity photon lifetime and a reduced average photon number. Consequently, the average photon number in the cavity decreases correspondingly with increasing dissipation. When the system is dissipationless, multiphoton processes gradually emerge over time under detuned conditions; however, the introduction of dissipation causes the system to decay downward before upward transitions can occur, thereby effectively suppressing the accumulation of multiphotons. Based on the above analysis, we need to choose an appropriate decay rate to balance this effect. 
      In the resonant case with $\delta = 0$ corresponding to Fig.~\ref{Fig5}(c), the black dashed curve shows that $G^{(2)}(0)=10^{-4}$   for ($g/\omega_{c}\sim 0.18,\gamma/\omega_{c} \sim  10^{-6}$) or  ($g/\omega_{c}\sim 0.8,\gamma/\omega_{c} \sim  4 \times 10^{-3}$), while the white dashed curve indicates that $G^{(2)}(0)=10^{-5}$ for ($g/\omega_{c} \sim 0.58,\gamma/\omega_{c} \sim  10^{-6}$) or ($g/\omega_{c} \sim 0.8,\gamma/\omega_{c} \sim 7 \times 10^{-4}$). For the detuned case of $\delta = 5\Omega_{p}$ in Fig.~\ref{Fig5}(d), the white dashed curve presents  $G^{(2)}(0)=10^{-3}$  for ($g/\omega_{c} \sim0.36, \gamma/\omega_{c}\sim  1.9\times10^{-4}$) or ($g/\omega_{c}\sim 0.6,\gamma/\omega_{c} \sim 10^{-6} $). 
      These results strongly demonstrate the realization of a single-photon source operating in the ultrastrong coupling regime. Specifically, for the parameter set $(g/\omega_{c}, \gamma/\omega_{c}) = (0.7, 10^{-6})$ in Fig.~\ref{Fig5}(c), the calculated $G^{(2)}(0)$ is $7.09 \times 10^{-6}$, which is further reduced to $3.32 \times 10^{-8}$ after normalization. For detuned driving with $\delta=5\Omega_{p}$, the same parameter point delivers a $G^{(2)}(0)$ of $7.31 \times 10^{-4}$, and the normalized result is $3.41 \times 10^{-5}$. To the best of our knowledge, previous investigations on cavity QED systems have only achieved a minimum $G^{(2)}(0)$ of $10^{-2}$. Notably, our optimized results are six orders of magnitude lower than those reported in prior studies~\cite{peng j,Source-1,Source-2}.
      
\begin{figure}[tbp]
      	\centering
      	\includegraphics[width=0.48\textwidth]{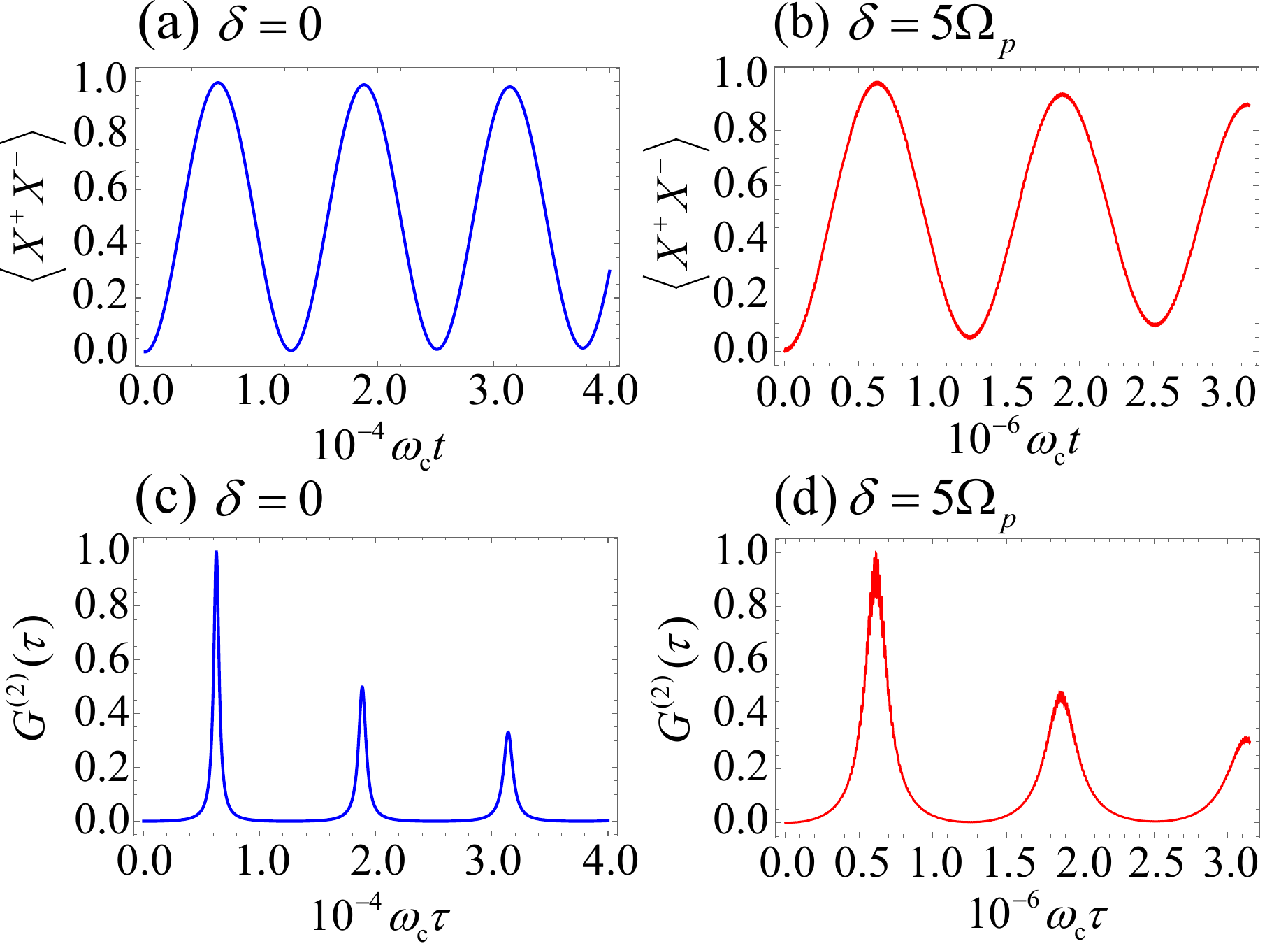}
      	\caption{ (a, b) Time evolution of the average photon number $\left\langle X^{+}X^{-}\right\rangle $ for
      		(a) resonant case ($\delta=0$) and (b) detuned case ($\delta=5\Omega_{p}$).
      		(c, d) Normalized second-order correlation function $G^{(2)}(\tau)$ vs $\tau$ for (c) resonant
      		case ($\delta=0$) and (d) detuned case ($\delta=5\Omega_{p}$). The common parameters are $\gamma=10^{-6}\omega_{c}$ and $g=0.7\omega_{c}$. Other parameters are the same
      		as in Fig.~\ref{Fig5}.}
      	\label{Fig8}
      \end{figure}
                                                    
      To be more clear, we further plot the time evolution of the average photon number $\langle X^{+} X^{-} \rangle$ and normalized $G^{(2)}(\tau)$ in Fig.~\ref{Fig8} by the same parameters $g/\omega_{c}$ and $\gamma/\omega_{c}$ discussed above,  namely,  $(g/\omega_{c},~\gamma/\omega_{c})=(0.7,~10^{-6})$.  
      As shown in Figs.~\ref{Fig8}(a) and \ref{Fig8}(b), the average cavity photon number $\left\langle X^{+}X^{-}\right\rangle $ oscillates periodically between 0 and 1 with a very slowly reducing amplitude. The oscillation period $T\approx1.26\times10^{-2}/\gamma $ in Fig.~\ref{Fig8}(a) and $T\approx0.13/\gamma $ in Fig.~\ref{Fig8}(b)  respectively.
       The values of  first and third peaks are 0.996 and 0.98 in Fig.~\ref{Fig8}(a), and  0.976 and 0.897 in  Fig.~\ref{Fig8}(b). The amplitudes of $\langle X^{+} X^{-} \rangle$ eventually reaches a steady value in a long enough time with a steady state of a superposition $|b,0\rangle$ and $|b,1\rangle$. This indicates that the system can continuously generate single photons under continuous-wave driving. However, in the time we concern, there are already enough peaks occur that indicate the generation of single-photons. This observation can be further verified in Figs.~\ref{Fig8}(c) and \ref{Fig8}(d), where $G^{(2)}(\tau)$ is normalized.  As shown in Figs.~\ref{Fig8}(c) and \ref{Fig8}(d),  $G^{(2)}(0)=3.32\times10^{-8}$ (Fig.~\ref{Fig8}(c)) and $3.41\times10^{-5}$ (Fig.~\ref{Fig8}(d) respectively, which are much less than 0.1, indicating a significant photon anti-bunching phenomenon. This is a clear characteristic of single photon emission. Moreover,  $G^{(2)}(\tau)$ periodically oscillates and exhibits much peaks with height reducing apparently. The locations of the peaks just match well with the peaks of $\langle X^{+} X^{-}\rangle$ in Figs.~\ref{Fig8}(a) and  \ref{Fig8}(b) respectively. Every time the photon number approaches its maximum, a single-photon emission occurs and $G^{(2)}(\tau)$ reaches maximum. Then, the system returns to the initial state $|b,0\rangle$. Under continuous driving the system transfers to $|b,1\rangle$ from $|b,0\rangle$ again and begins another photon emission cycle.  The above processes repeats periodically, demonstrating periodic single-photon release. We also note that the summation of the probabilities on $|b,0\rangle$ and $|b,1\rangle$ is almost 1 during the above process.    
       \begin{figure}[t!]
        \centering
      \includegraphics[width=0.48\textwidth]{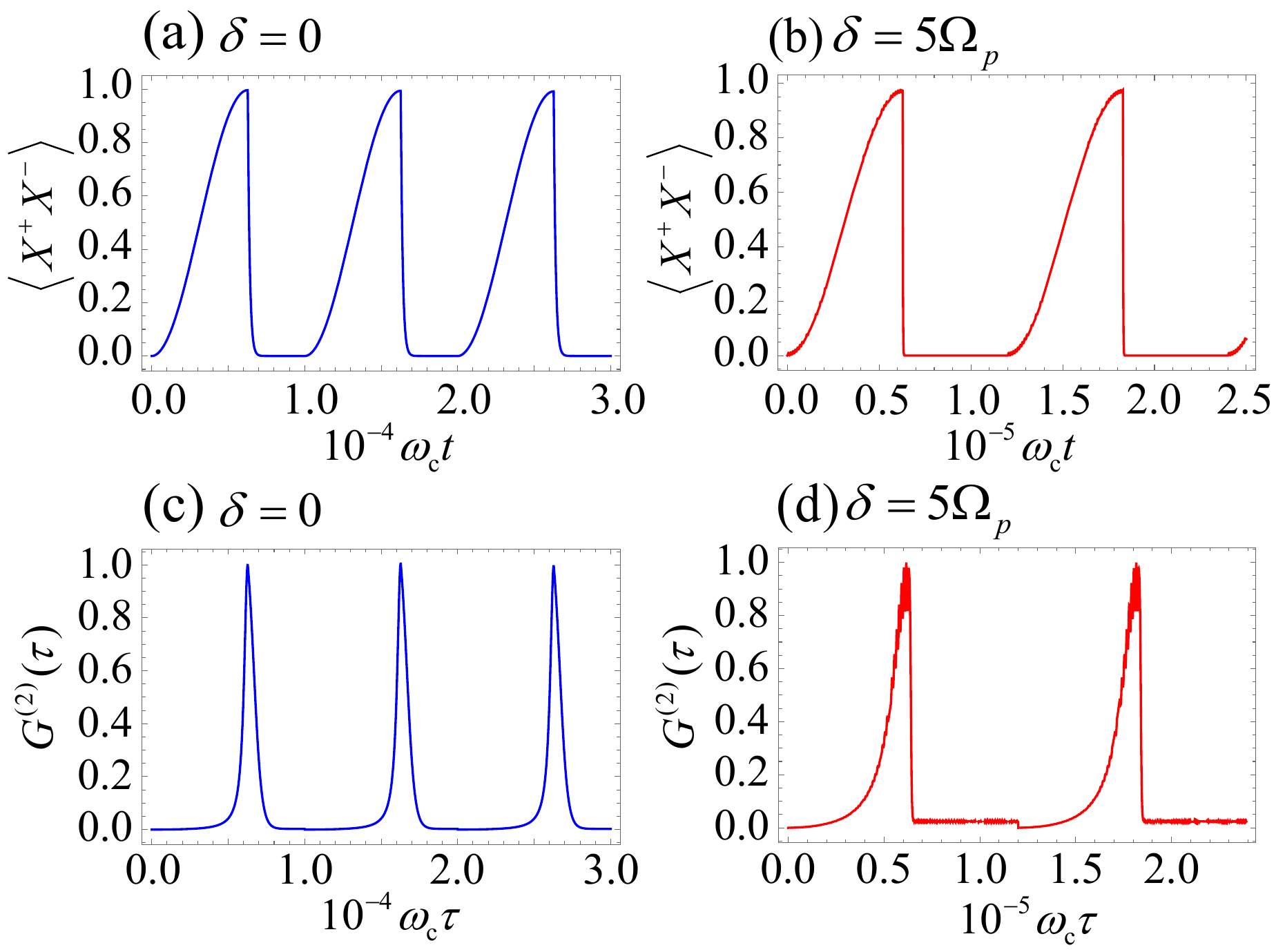}
      	\caption{Evolution of $\left\langle X^{+}X^{-}\right\rangle$ and $G^{(2)}(\tau)$  with the driving fields $\Omega_{p,s}$ periodically turned on at $t=n T/\omega_{c}$ ($n=0,1,2,3,...$) and off at the peaks of $\left\langle X^{+}X^{-}\right\rangle$, accompanied by switching the cavity decay rate $\gamma_{3}$  from $\gamma_{0}=10^{-6}\omega_{c}$ to $\gamma'=8\times10^{-3}\omega_{c}$ simultaneously. (a, c)  Resonance case ($\delta=0$) with periodicity $T=10^{4}/\omega_{c}$;
(b, d) Detuned case ($\delta=5\Omega_{p}$) with periodicity  $T=1.2\times10^{5}/\omega_{c}$. Other parameters are the same
      		as in Fig.~\ref{Fig8}.}
      	\label{Fig6.0}
      \end{figure}
      To evaluate the performance of the single photon source, we further analyze the indistinguishability $I$ of the emitted photons.  Here $I$ is obtained by the similar formula in the strong coupling regime but replacing $g^{(2)}(\tau)$ by $G^{(2)}(\tau)$. 
      As shown in Figs.~\ref{Fig8}(c) and \ref{Fig8}(d), the indistinguishability of the emitted photons is 99.10\% and 96.10\% respectively, which is close to 100\%. Our scheme works well for both resonant and detuned case in the ultrastrong coupling regime.

     To collect single photons, we apply the same pulse driving as in Eq. (\ref{kt}), but replace the cavity decay rate $\kappa_{3}(t), \kappa_{0},\kappa^{'}$ by $\gamma_{3}(t), \gamma_{0}$ and $\gamma^{'}$.  $t_{1}$  is the moment of the first peak of  $\langle X^{+} X^{-} \rangle$.
Using the same parameters as in Fig.~\ref{Fig8}, we numerically simulate $\langle X^{+} X^{-} \rangle$ and $G^{(2)}(\tau)$ for  $T=10^{4}/\omega_{c} $ (resonant) and  $T=1.2 \times 10^{5}/\omega_{c} $ (detuned) and show the results in Fig.~\ref{Fig6.0}. 
     As shown in Figs.~\ref{Fig6.0}(a) and \ref{Fig6.0}(b), $\langle X^{+} X^{-} \rangle$ periodically oscillates between $0$ and $1$ for both resonant and detuned cases.
The normalized  $G^{(2)}(\tau)$ [Figs.~\ref{Fig6.0}(c) and \ref{Fig6.0}(d)] shares the same period as $\langle X^{+} X^{-} \rangle$, featuring peaks with the identical height. $G^{(2)}(0)=3.32\times10^{-8}$ and $G^{(2)}(0)=3.41 \times 10^{-5}$, which is approaching zero. This is clear evidence of single photon source. To evaluate the performance of the single photon source, we explore the single-photon efficiency, defined as
      \begin{equation}
      \epsilon =\gamma_{3} \int \langle X^{+}(t) X^{-}(t) \rangle dt. 
      \end{equation}
 $\epsilon$ reaches 99.96\% and 100\% for the resonant and detuned cases, respectively. 
 The corresponding indistinguishability $I$ are 98.98\%  (resonant) and 95.91\% (detuned), respectively. The purity $P$ remains the same as the continuous-wave driving case, namely $P=99.99\%$ (resonant) and $P=99.93\%$  (detuned).  Therefore, the system exhibits excellent performance for single-photon source in the ultrastrong coupling regime.

\section{Discussion and Conclusion }~\label{conclusion}   
We discuss the experimental feasibility of the proposed scheme, whose key experimental prerequisites are the realization of the ultrastrong Jaynes–Cummings (JC) model and the implementation of a $\triangle$-type atomic configuration. Notably, our scheme is applicable to both strong and ultrastrong coupling regimes, which significantly enhances its experimental flexibility.
The ultrastrong coupling regime has been experimentally realized on multiple state-of-the-art platforms, including superconducting circuits~\cite{us-super1,us-super2,us-super3,deep-1}, Landau polaritons~\cite{us-ILLT1,us-ILLT2,us-ILLT3,Bayer2017}, organic molecules~\cite{us-mole1,us-mole2,us-mole3}, optomechanical systems~\cite{opto-1}, and intersubband polaritons~\cite{us-isbt1,us-isbt2,us-isbt4}. Impressively, deep-strong coupling with $g/\omega_{c}=1.34$ and $g/\omega_{c}=1.43$ has been reported in circuit QED systems and semiconductor quantum wells, respectively~\cite{deep-1,Bayer2017}. The effective ultrastrong JC model can be derived from the full quantum Rabi model by dynamically suppressing counter-rotating terms via external field modulation~\cite{JC-2,JC-3,Huang2017PRA}. In addition, exact JC-type coupling can be naturally established when the two upper atomic levels feature an angular momentum difference of $\Delta m=\pm1$ and interact with circularly polarized optical fields~\cite{JC-1}.
Regarding the $\triangle$-type cyclic atomic configuration, such closed-transition structures are absent in natural atoms due to inherent selection rules. Nevertheless, they can be faithfully implemented in symmetry-broken quantum systems, such as chiral molecules~\cite{chiral-1,chiral-2,artifical symmetry-broken-1} and artificially engineered quantum atoms with broken symmetry~\cite{artifical symmetry-broken-2}. In particular, the $\triangle$-type artificial atom has been experimentally demonstrated in superconducting quantum circuits~\cite{artificial-2}.
Furthermore, the dissipative parameters adopted in our model are also experimentally accessible. The cavity decay rate $\kappa/\omega_{c}$ and atomic decay rate $\gamma/\omega_{c}$ ranging from $10^{-6}$ to$10^{-3}$ have been achieved in various cavity QED~\cite{three c,Smith2025PRAP} and circuit QED platforms~\cite{Schoelkopf2011PRL,Sacepe2025}. Moreover, dynamical tuning of the cavity decay rate has been experimentally validated in previous works~\cite{kappa}.
Collectively, all the required physical conditions and key parameters of our scheme are achievable with current experimental technologies, confirming the practical experimental accessibility of the proposed deterministic single-photon source.

\begin{figure}[b!]	
	\centering
	\includegraphics[width=0.48\textwidth]{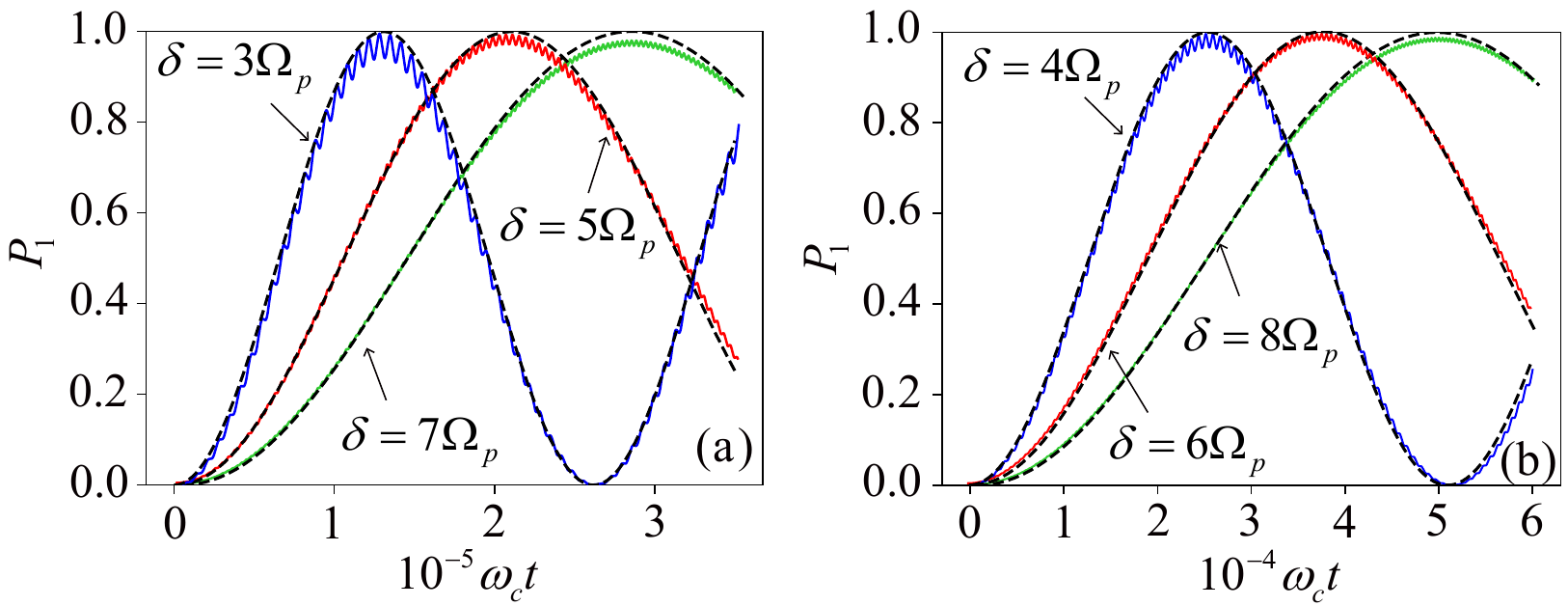}
	\caption{ (Color online) Population $P_{1}$ of state $|b,1\rangle$ versus time, obtained from the effective Hamiltonian (\ref{eq:A7}) (black dashed lines) and  the exact Hamiltonian (\ref{eq:6})  (solid lines) for multiple detunings in the (a) strong-coupling  and (b) ultrastong-coupling regimes. Parameters: (a) $g = 0.06\omega_{c}$, $\Omega_{p} = 3 \times 10^{-4}\omega_{c}$ , $\delta = 3\Omega_{p}$ (blue), $5\Omega_{p}$ (red), and $7\Omega_{p}$ (green); (b) $g = 0.5\omega_{c}$, $\Omega_{p} = 2 \times 10^{-3}\omega_{c}$, $\delta= 4\Omega_{p}$ (blue), $6\Omega_{p}$ (red), and $8\Omega_{p}$ (green). Common parameters: $\Omega_{s}=\alpha\Omega_{p},~\omega_{s}=\omega_{p}-\omega_{c},~\omega_{b} = -5\omega_{c},~\omega_{m}=-0.5\omega_{c}$, and $\omega_{e}=0.5\omega_{c}$.  All parameters are given in units of $\omega_{c}$.}
	\label{Fig2}
\end{figure}

We propose a practical scheme for a deterministic single-photon source based on a $\triangle$-type three-level atom coupled to a single-mode cavity field. By applying two classical driving fields, deterministic single-photon emission can be generated in both strong and ultrastrong coupling regimes. For continuous-wave driving, the system exhibits excellent single-photon performance. In the strong coupling regime, the normalized equal-time second-order correlation function is $g^{(2)}(0)\sim10^{-6}$, with a photon indistinguishability of $98.73\%$ and a state purity of $99.95\%$. In the ultrastrong coupling regime, the corresponding normalized equal-time second-order correlation function is $G^{(2)}(0)\sim10^{-8}$, achieving an indistinguishability of $99.10\%$ and a purity of $99.99\%$. For pulsed driving within the ultrastrong coupling regime, the emission efficiency, indistinguishability, and purity reach $99.96\%$, $98.98\%$, and $99.99\%$ under resonant conditions, and $100\%$, $95.91\%$, and $99.93\%$ under detuned conditions, respectively. The near-perfect performance of the proposed single-photon source addresses the long-standing challenge in linear optical quantum computing. This work provides a feasible route for the implementation of deterministic single-photon sources, promotes the advancement of quantum information science, and deepens the physical understanding of fundamental quantum mechanisms.

\begin{acknowledgments}
	We gratefully acknowledge Jie Peng for technical support. J.-F. H. is partially supported by the National Natural Science Foundation of China (Grant No. 12475016) and the Science and Technology Innovation Program of Hunan Province (Grant No. 2025RC3141).
\end{acknowledgments}

\appendix
\section{Validity of the effective Hamiltonian  (\ref{eq:A7}) and (\ref{eq:A1})}

To show the validity of the effective Hamiltonian (\ref{eq:A7}) and (\ref{eq:A1}),  we calculate the probability  $P_{1}$ in state $|b,1\rangle$ from initial state $|b,0\rangle$ for strong and ultrastrong coupling regimes, respectively.  Then we compare the results with that numerically obtained from the exact Hamiltonian (\ref{eq:6}) for large detuning and resonant cases respectively.

\subsection{Large detuning case}

To verify the validity of the effective Hamiltonian (\ref{eq:A7}), we 
calculate the probability  $P_{1}$ from both effective Hamiltonian (\ref{eq:A7}) and exact Hamiltonian~(\ref{eq:6}) with the large detuning condition $\varepsilon_{1}-\omega_{b}-\omega_{p}\gg\chi_{p}$ and $\varepsilon_{1}-\omega_{b}-\omega_{c}-\omega_{s}\gg\chi_{s}$.

\begin{figure}[tbp]
\centering
	\includegraphics[width=0.48\textwidth]{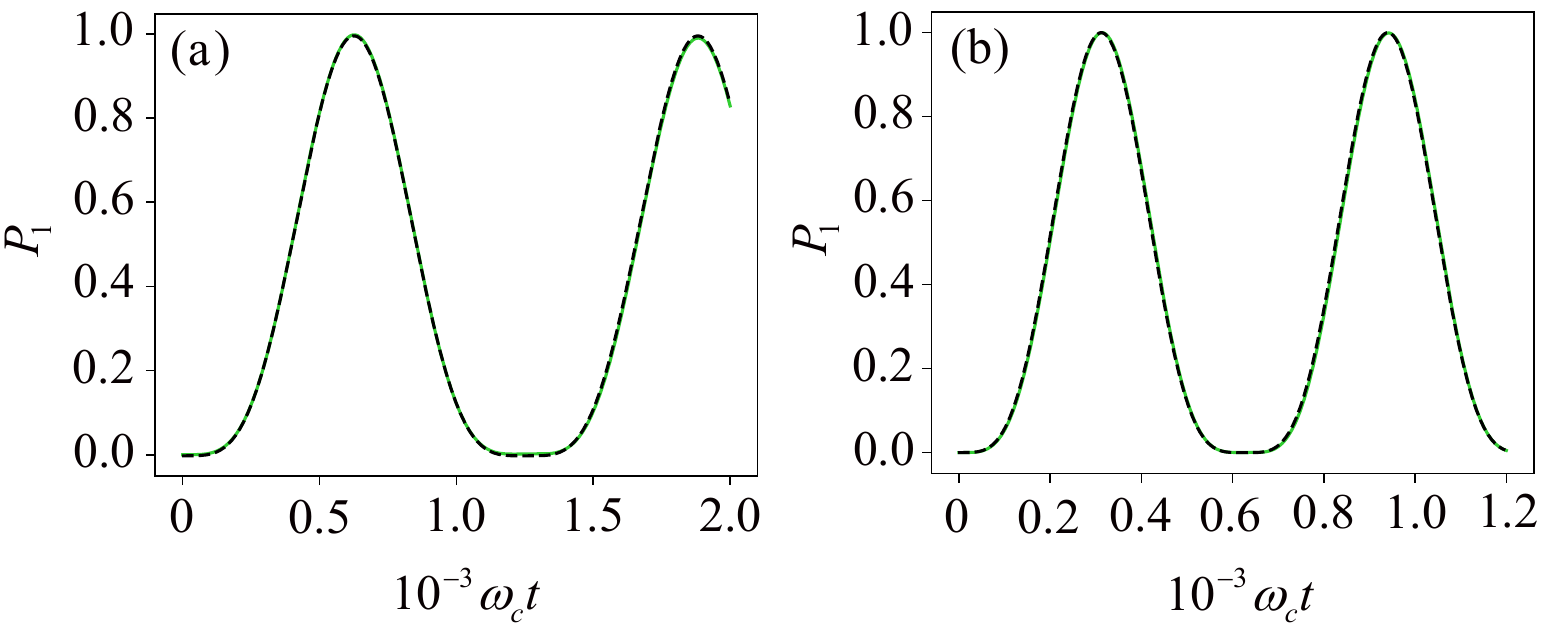}
	\caption{(Color online) Time evolution of the single-photon occupation probability $P_{1}$ of state $|b,1\rangle$ at resonance ($\delta=0$). Black dashed curves: effective Hamiltonian Eq. (\ref{eq:A1}); green solid curves: full numerical simulations of the exact Hamiltonian Eq.  (\ref{eq:6}). Results are shown for (a) strong-coupling and (b) ultrastrong-coupling regimes.  (a)  $g = 0.05\omega_{c}$, $\Omega_{p} = 10^{-3}\omega_{c}$ and  $\alpha=\alpha_{c}$. (b)  $g = 0.5\omega_{c}$, $\Omega_{p} = 2 \times 10^{-3}\omega_{c}$ and  $\alpha=\alpha_{c}$. The rest of the parameters are the same as Fig.~\ref{Fig2}.}
	\label{Fig2-1}
\end{figure}  

Fig.~\ref{Fig2} presents the analytical and numerical results in both strong ($g/\omega_{c}=0.06$) and ultrastrong ($g/\omega_{c}=0.5$)  coupling regimes. By varying the detuning $\delta$, the  numerical results were
compared with the predictions of the effective Hamiltonian (\ref{eq:A7}), as depicted in Fig.~\ref{Fig2}.  Figure~\ref{Fig2}(a) shows that a complete Rabi oscillations predicted by the effective Hamiltonian (\ref{eq:A7}) are in good agreement with the numerical results in strong coupling regime. As the detuning increases, a 4\% deviation is observed at $\delta=7\Omega_{p}$. This deviation arises because an increase in detuning reduces
 the difference between the effective coupling strengths $g_{0,1}$ and $g_{1,2}$, which in turn weakens the system's ability to suppress multiphoton transitions. When the coupling strength enters the ultrastrong coupling region, as shown in Fig.~\ref{Fig2}(b), the analytical and numerical solutions are highly consistent, achieving perfect sate transfer between the states $|b,0\rangle$ and $|b,1\rangle$. This phenomenon verify the physical mechanism to generate single photons since the atom state $|b\rangle$ is decoupled from the cavity field. The results demonstrates that the exact Hamiltonian (\ref{eq:6})  can be well approximated by the effective Hamiltonian  (\ref{eq:A7}), which capture the main mechanism to generate single photons.

\subsection{Resonance case}
We consider the resonant case: $\varepsilon_{1}-\omega_{b}-\omega_{p}=0$ and $\varepsilon_{1}-\omega_{b}-\omega_{c}-\omega_{s}=0$. In order to verify the validity of the effective three-level Hamiltonian~(\ref{eq:A1}) , we numerically calculate the evolution of $P_1$ governed by Hamiltonian~(\ref{eq:6}) from $|b,0\rangle$ for $g = 0.05\omega_{c}$   and $g = 0.5\omega_{c}$ respectively.  Then we compared the exact results with analytical result Eq. (\ref{eq:A2}) obtained from the effective Hamiltonian~(\ref{eq:A1})  in Fig.~\ref{Fig2-1}. 
Our results indicate that, the Rabi oscillation predicted by the effective Hamiltonian Eq.~(\ref{eq:A1}) is in well agreement with that predicted by the exact Hamiltonian~(\ref{eq:6}) both in strong and ultrastrong coupling regimes. Moreover, when the ratio $\alpha $ between the amplitudes of the driving fields reaches the optimal value $\alpha = \alpha_{c}$, the peak value of the probability $P_{1}$ can reach 1, which confirms the perfect single-photon generation.

\end{document}